\documentclass[prl,superscriptaddress,twocolumn]{revtex4-2}
\usepackage[dvipsnames]{xcolor}
\usepackage[english]{babel}
\usepackage{amsmath,amsfonts,amssymb}
\usepackage{graphicx,float,calc}
\usepackage[breaklinks,colorlinks,urlcolor=blue,citecolor=blue,linkcolor=blue]{hyperref}
\usepackage{soul}
\usepackage{lineno}  
\sethlcolor{red}
\soulregister{\ref}7
\begin{document}
 \title{Fast and Robust Remote Two-Qubit Gates on Distributed Qubits
 }
	
	\author{Yunan Li}\thanks{These authors contributed equally to this work.}
	\affiliation{National Laboratory of Solid State Microstructures, School of Physics,
		Nanjing University, Nanjing 210093, China}
	
	\affiliation{Shishan Laboratory, Suzhou Campus of Nanjing University, Suzhou 215000, China}
	
	\author{Xi Zhang}\thanks{These authors contributed equally to this work.}
	\affiliation{National Laboratory of Solid State Microstructures, School of Physics,
		Nanjing University, Nanjing 210093, China}
	
	\affiliation{Shishan Laboratory, Suzhou Campus of Nanjing University, Suzhou 215000, China}	
  \author{Weixin Zhang}\thanks{These authors contributed equally to this work.}
	\affiliation{Key Laboratory of Atomic and Subatomic Structure and Quantum Control (Ministry of Education), Guangdong Basic Research Center of Excellence for Structure and Fundamental Interactions of Matter, and School of Physics, South China Normal University, Guangzhou 510006, China}

	\author{Ruonan Guo}
	\affiliation{National Laboratory of Solid State Microstructures, School of Physics,
		Nanjing University, Nanjing 210093, China}
		\affiliation{Shishan Laboratory, Suzhou Campus of Nanjing University, Suzhou 215000, China}
	
     \author{Yu Zhang}
	\affiliation{National Laboratory of Solid State Microstructures, School of Physics,
		Nanjing University, Nanjing 210093, China}
	
	\affiliation{Shishan Laboratory, Suzhou Campus of Nanjing University, Suzhou 215000, China}	

	\author{Xinsheng Tan}
	\email{tanxs@nju.edu.cn}
	\affiliation{National Laboratory of Solid State Microstructures, School of Physics,
		Nanjing University, Nanjing 210093, China}
		\affiliation{Shishan Laboratory, Suzhou Campus of Nanjing University, Suzhou 215000, China}
 	\affiliation{Synergetic Innovation Center of Quantum Information and Quantum Physics, University of Science and Technology of China, Hefei, Anhui 230026, China}
  	\affiliation{Hefei National Laboratory, Hefei 230088, China}

	\author{Yang Yu}
         \email{yuyang@nju.edu.cn}
	\affiliation{National Laboratory of Solid State Microstructures, School of Physics,
		Nanjing University, Nanjing 210093, China}
		\affiliation{Shishan Laboratory, Suzhou Campus of Nanjing University, Suzhou 215000, China}
     	\affiliation{Synergetic Innovation Center of Quantum Information and Quantum Physics, University of Science and Technology of China, Hefei, Anhui 230026, China}
          \affiliation{Hefei National Laboratory, Hefei 230088, China}

	\begin{abstract}    
		{
        Distributed quantum computing offers a potential solution to the complexity of superconducting chip hardware layouts and error correction algorithms. High-quality gates between distributed chips enable the simplification of existing error correction algorithms. This article proposes and demonstrates a remote quantum geometric gate scheme via parametric modulation. Our scheme inherits the intrinsic robustness of geometric phases. Meanwhile, by employing gradient-based optimization algorithms(Adaptive Moment Estimation) from deep learning, we design control waveforms that significantly suppress population leakage. We experimentally realize the rapid remote SWAP and $\sqrt{\text{SWAP}}$ gates with high fidelity, completing operation in about 30 ns. The gate error of SWAP ($\sqrt{\text{SWAP}}$) is 1.16\% (0.91\%) after excluding the effect of energy relaxation. The simulation  demonstrate that this scheme can be implemented in the distributed chips connected by cables extending several meters. Our results highlight the effectiveness of the proposed protocol in enabling modular quantum processors, offering a promising path toward the realization of fault-tolerant quantum computation.
		}    
	\end{abstract}

	\maketitle
 
 Quantum computation has demonstrated significant application potential across multiple domains over the past few decades\cite{Feynman_1982,DiVincenzo_2000,Nielsen_Chuang_2010}.
 The realization of fault-tolerant quantum computation, recognized as a critical near-term objective, has witnessed remarkable progress through both theoretical and experimental advancements\cite{GoogleQuantumAIAndCollaborators_2025,GoogleQuantumAI_2023,Krinner_2022}.
  The superconducting circuit is considered a promising candidate for implementing fault-tolerant architectures, owing to its exceptional scalability and well-established control protocols\cite{Koch_2007,Wang_2022,Barends_2013,Gyenis_2021,Hyyppa_2022,Kalashnikov_2020,Manucharyan_2009,Yan_2016}.
  Nevertheless, a fundamental limitation of current superconducting quantum processors lies in their reliance on nearest-neighbor coupling configurations\cite{Gao_2025,Arute_2019_all}.
  This architectural constraint substantially increases the operational overhead of quantum error correction protocols\cite{Bravyi_2010}, imposing severe restrictions on system scalability.

  Recent developments in remote connectivity schemes offer promising solutions to these challenges. Theoretical studies have established that non-local coupling architectures can reduce the complexity of quantum error correction algorithms, such as quantum low-density parity-check (LDPC) codes\cite{Gottesman_2014, Baspin_2022b, Breuckmann_2021, Baspin_2022}. Some long-distance coupling LDPC schemes have been demonstrated\cite{Krishna_2021a,Cohen_2022,Xu_2022,Xu_2024,Bravyi_2024}, which theoretically extend the fault-tolerant capability of the superconducting platform. Notably, remote connectivity plays a critical role at the hardware level. Remote connectivity structures effectively address challenges such as heat dissipation and insufficient space in a dilution refrigerator\cite{Krinner_2019,Magnard_2020}. Furthermore, they mitigate on-chip frequency crowding while enhancing the scalability of quantum processors, establishing a viable framework for implementing large-scale distributed quantum computing\cite{Krinner_2019,Saida_2024,Stassi_2020a}, quantum internet protocols\cite{Kimble_2008,Wehner_2018}, and hybrid quantum computing\cite{Xiang_2013}. Remarkably, the remote connectivity platform enables groundbreaking investigations into fundamental quantum phenomena, particularly long-distance entanglement generation and Bell's inequality tests\cite{Gold_2021}.
  
\begin{figure*}[!ht]
    \centering
    \includegraphics[width=18cm]{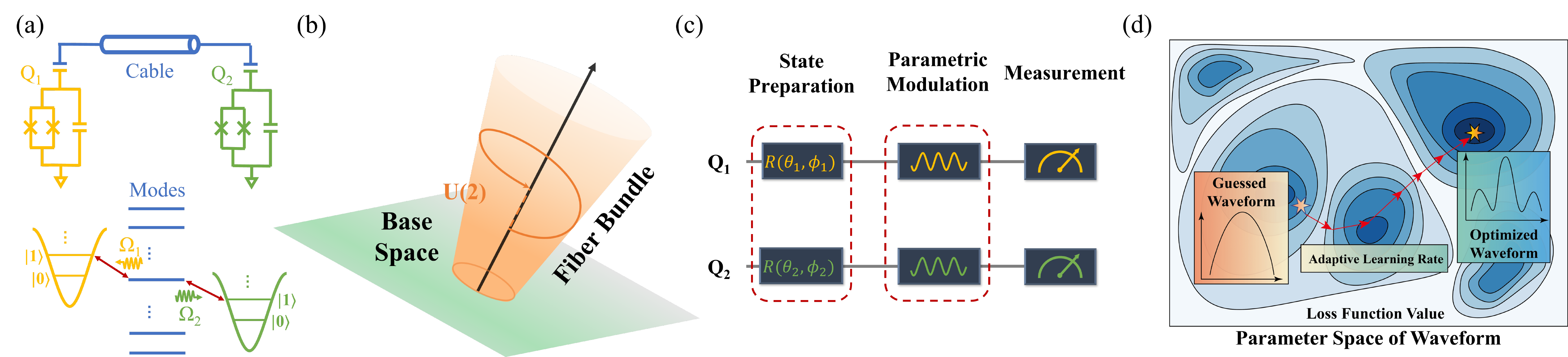}
    \caption{{\bf Demonstration of Remote Connectivity Scheme.} (a) Sample structure. Qubits were capacitively coupled to the 15cm Al cable. In the actual device, the qubits are connected to cables via couplers. In this experiment, the coupler is treated as a capacitor. Therefore, for clarity, it is omitted from the schematic diagram. We applied parametric modulation on two qubits and constructed coupling between qubits and cable modes. (b) Holonomic Evolution. Cyclic evolution in the base space induces non-Abelian geometric phases in the U(2) bundle. (c) Timing sequence of our experiment. The initial state was built by two arbitrary driving pulses on $Q_1$ and $Q_2$. Two parametric modulation pulses were simultaneously applied on $Q_1$ and $Q_2$ while realizing holonomic evolution. (d) Demonstration of Adam Optimizer. Given a guessed waveform, the Adam optimizer leverages its adaptive learning rate to avoid local optima, enabling the loss function to converge to the global optimum and effectively suppress cable mode leakage.}
    \label{fig:FiG1}
\end{figure*}

All cable modes are involved in long-distance connectivities (typically on the order of several meters) due to the small free spectral range (FSR). In such scenarios, state transfer predominantly relies on itinerant photon schemes\cite{Kurpiers_2018,Zhong_2019,Qiu_2023,Storz_2023}, demanding precise control of coupling strength and high-quality factor cables. In contrast, the larger FSR enables one-mode protocol for short-distance connections to achieve state transfers, as exemplified by the modulation of flux bias and microwaves\cite {Chang_2020,Zhong_2019, Leung_2019,Malekakhlagh_2024,Mollenhauer_2024, Song_2024, Zhong_2021,Niu_2023}. These schemes, involving dynamic pulses and adiabatic evolution, exhibit distinct trade-offs. Developing rapid and fault-tolerant methods remains crucial for implementing universal quantum gates on distributed quantum chips. 

This article proposes and experimentally demonstrates a remote two-qubit gate protocol on distributed qubits.
We employ the cable modes as auxiliary energy levels to realize the holonomic gates. This control strategy retains the intrinsic robustness of geometric phase, a property that has been extensively validated across multiple physical platforms\cite{AbdumalikovJr_2013,Egger_2019,Xu_2018,Yan_2019,Zhang_2019,Feng_2013,Arroyo-Camejo_2014,Nagata_2018,Sekiguchi_2017,Zhou_2017,Zu_2014,Ai_2020}. In contrast to previous schemes, the narrow spacing between cable modes renders the system vulnerable to population leakage. To mitigate this effect, we perform parameter optimization using the adaptive moment estimation (Adam) method, the gradient-based algorithms widely used in deep learning.  This approach leads to a significant suppression of leakage and a marked reduction in the overall error rate.

To enhance the dimensionality of the control parameter space, we utilize  parametric modulation to realize the couplings between superconducting qubits and the coaxial cable. Our implementation achieves SWAP and $\sqrt{\text{SWAP}}$ gate durations of approximately 30 ns with a gate error of approximately 1\%, representing a significant improvement over existing approaches. We use numerical simulations to select the optimal frequency parameters and waveform shape to suppress critical population leakage, while the results demonstrate refrigerator-level cable-length extension capabilities. Our results demonstrate the potential of this approach to effectively suppress cable mode leakage at small FSR values. This architecture provides a viable pathway toward fault-tolerant quantum computation in distributed systems.

\textbf{Remote Connectivity Qubits.} In our framework, two Transmon qubits were separately placed in two sample carriers, and these two qubits were capacitively coupled to the same 15cm Al coaxial cable\cite{supp}, corresponding to the FSR of 403 MHz, as shown in Fig.\ref{fig:FiG1}(a). 
  The coaxial cable functions as a multimode resonator, where each mode directly couples to the qubits with coupling strength $g_{ij}$ ($i=1,2$ for qubits, $j=0,1,2,...$ for resonator modes).
  Notably, all $g_{1j}$ parameters are positive, and for $g_{2j}$, positive and negative values alternate with different $j$\cite{Malekakhlagh_2024}. 
  The system Hamiltonian is given by:
 	\begin{equation}
		\begin{split}
			H/{\hbar}&=\sum_{i=1,2;j=0,1,2,...,n} \left[\omega_{Q_i}a^{\dagger}_{i}a_{i} + \omega_{M_j}b^{\dagger}_{j}b_{j}\right.\\
      &\left.+\frac{\alpha_i}{2}a_i^{\dagger}a_i^{\dagger}a_ia_i +(g_{ij}a_{i}^+b_{j} + \text{H.c.})\right] ,
		\end{split}
	\end{equation}
with flux-tuned qubit frequencies $\omega_{Q_1}/{2\pi} = 6.127 \text{GHz}$ and $\omega_{Q_2}/{2\pi} = 5.712 \text{GHz}$. The $\alpha_i/{2\pi} = -162\text{MHz}$ refers to the anharmonicities of these two qubits. $a_i^{\dagger} (a_i)$ is the associated creation (annihilation) operator of the qubits truncated to the lowest three levels. We selectively consider cable modes which are close to the qubit frequency, where we choose $\omega_{M_1}/{2\pi} = 6.36\text{GHz}$,\ $\omega_{M_2}/{2\pi} = 5.83\text{GHz}$,\ $\omega_{M_3}/{2\pi} = 5.38\text{GHz}$. $b_i^{\dagger} (b_i)$ is the associated creation (annihilation) operator of the cable modes truncated to the lowest two levels. Direct coupling strengths were measured, where $g_{12}/2\pi=30.26\text{MHz}$ and $g_{22}/2\pi=26.88\text{MHz}$.

\begin{figure*}[!ht]
    \centering
    \includegraphics[width=18cm]{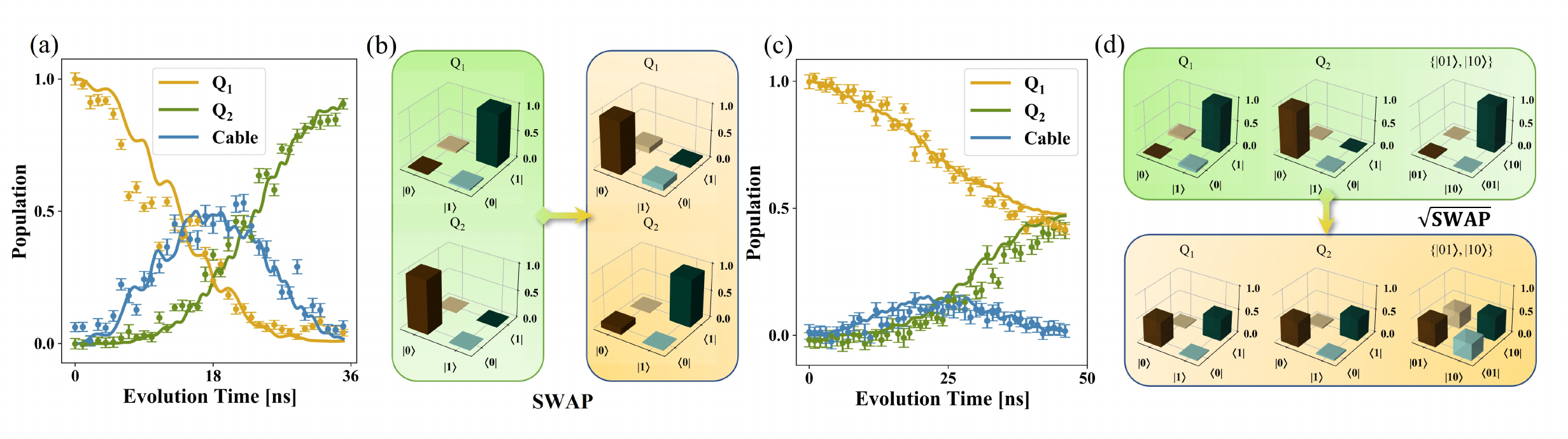}
    \caption{{\bf Remote Holonomic SWAP and $\sqrt{\text{SWAP}}$ Gate.} (a)(c) The population dynamics of the remote holonomic SWAP and $\sqrt{\text{SWAP}}$ gates. Orange, green and blue solid lines refer to the numerical results of $Q_1$, $Q_2$ and cable mode $M_2$ population, and dots refer to the experimental results. (b) Qubit tomography result of SWAP gate, the final state fidelity achieved $99.91\pm0.06\%$ considering the decoherence. (d) Subspace tomography result of $\sqrt{\text{SWAP}}$ gate. For single qubit tomography, $Q_2$ finally received 47.48\% population. For subspace tomography, we consider the subspace $\{|10\rangle,|01\rangle\}$. The final state fidelity achieved $97.54\pm0.42\%$, considering the decoherence.}
    \label{fig:FiG2}
\end{figure*}

  \textbf{Holonomic Gates via Parametric Modulation.} The geometric interpretation of holonomic gates is schematically depicted in Fig.\ref{fig:FiG1}(b),
  where cyclic evolution in the base space induces non-Abelian geometric phases in the U(2) bundle\cite{AbdumalikovJr_2013,Zu_2014}.
  In our approach, we implement parametric modulation\cite{caldwellParametricallyActivatedEntangling2018,chuScalableAlgorithmSimplification2023,hanErrorAnalysisSuppression2020,mckayUniversalGateFixedFrequency2016,seteParametricResonanceEntanglingGates2021,yanTunableCouplingScheme2018} to realize time-dependent couplings.
  Two independent parametric modulation flux pulses are applied to two qubits, generating the Hamiltonian form of $H_L(t)/\hbar= \sum_{i=1,2} A_i(t)\cos(\Delta_i t)a^{\dagger}_ia_i$ with flux pulse amplitude $A_i$ and modulation frequency $\Delta_i = \omega_{Q_i} - \omega_{M_2} + \delta_i$, where $\delta_i$ arises from frequency shift during parametric modulation.
  The system Hamiltonian can be written as $H(t)=H_0+H_L(t)$. In the rotating frame, the effective Hamiltonian becomes $H(t) = \tilde{g}_{12}(t)a^{\dagger}_{1}b_2 + \tilde{g}_{22}(t)a^{\dagger}_{2}b_2 + \text{H.c.}$, where $\tilde{g}_{i2}$ denotes the modulation-induced effective coupling strength.
  For universal gate operation in the $\{|10\rangle, |01\rangle\}$ subspace, we design cyclic evolution satisfying $\int_0^T\tilde{g}_{eff}(t)\ \text{d}t=\pi$, where effective holonomic coupling strength $\tilde{g}_{eff}(t)=\sqrt{\tilde{g}_{12}(t)^2+\tilde{g}_{22}(t)^2}$. This yields the unitary transformation:

  \begin{equation}
		\begin{aligned}
			U = 
			\begin{bmatrix}
				\cos\theta & e^{i\phi}\sin\theta \\
				e^{-i\phi}\sin\theta & -\cos\theta
			\end{bmatrix},
		\end{aligned}
	\end{equation}
  with $-e^{i\phi}\tan\frac{\theta}{2}=\tilde{g}_{12}/\tilde{g}_{22}$. 
  
Apparently we can control the strength of $\tilde{g}_{12}$ and $\tilde{g}_{22}$ to implement various quantum gate. 
Notably, the holonomic gate can be theoretically implemented within significant short time scales. 
 The tunability of parametric modulation not only ensures gate universality but also enables precise temporal control of holonomic operations, which may help us overcome the leakage and decoherence problem.
According to the parameters of our experimental sample, the Adam algorithm provided several possible options. In the experiment, we employed a time-dependent waveform with smooth rising and falling edges to reduce gate control errors.

\textbf{State Transfer and Entanglement.} In this article, we demonstrate the remote ${\text{SWAP}}$  and $\sqrt{\text{SWAP}}$ gates, which are the elementary operations for realizing universal quantum circuits. Fig.\ref{fig:FiG1}(c) shows the timing sequence of our experiments with the system initialized to $|10\rangle$. By varying the value of $\tilde{g}_{12}/\tilde{g}_{22}$, general two-qubit gates can be constructed within this holonomic architecture. In our experiments, single parametric modulation was first calibrated. We idled the qubit at its hillside frequency and mainly considered its first-order parametric modulation coupling. Through precise adjustment of flux pulse amplitude $A_i$, the $A_i-\delta_i$ relation was experimentally established. To determine the $\delta_i$, simultaneous flux pulses were  applied on both qubits. 
Using the designed waveform, we extract the average envelope $g^a_{i2}$, and optimize the gate duration as $t = \pi/g^a_{eff}$, with $g^a_{eff} = \sqrt{{g^a_{12}}^2+{g^a_{22}}^2}$. 
To characterize the gate operation, quantum state tomography was performed on our system, facilitating the optimization of $\delta_i$ parameters\cite{supp}.

  Quantum state transfer between distinct quantum network nodes can be achieved via SWAP gate. As shown in Fig.\ref{fig:FiG2}(a) population dynamics of $Q_1$, $Q_2$ and $M_2$ were experimentally observed. Here we set $\tilde{g}_{12}(t)=\tilde{g}_{22}(t)$, while the envelope average $g^a_{12}/2\pi=g^a_{22}/2\pi = 10.10\text{MHz}$. The desired pulse were engineered between both qubits and cable mode $M_2$, resulting in a gate duration of 35 ns. Theoretically, we can achieve faster gate operation on this sample, but experimentally, we have problems achieving greater coupling (probably due to the influence of the unwanted two-level systems).
  During the transferring process, a partial population was transferred to $M_2$ and then transferred back. Fig.\ref{fig:FiG2}(b) presented the quantum state tomography results of two distributed qubits. Density matrices of $Q_1$ and $Q_2$ were reconstructed at the initial and final states. Final state tomography on $Q_2$ yielded a fidelity quantified by $\mathcal{F}=Tr(\sqrt{\sqrt{\rho}\sigma\sqrt{\rho}})$ reaching $99.91\pm0.06\%$. Here, $\rho$ refers to the experimental result of our tomography, and $\sigma$ refers to the simulation result after considering the decoherence effect, indicating that the experimental error is basically due to decoherence. The rest of the dominant error source was identified as residual cable mode leakage, which will be systematically discussed in the methods part.
  
\begin{figure*}[!ht]
    \centering
    \includegraphics[width=18cm]{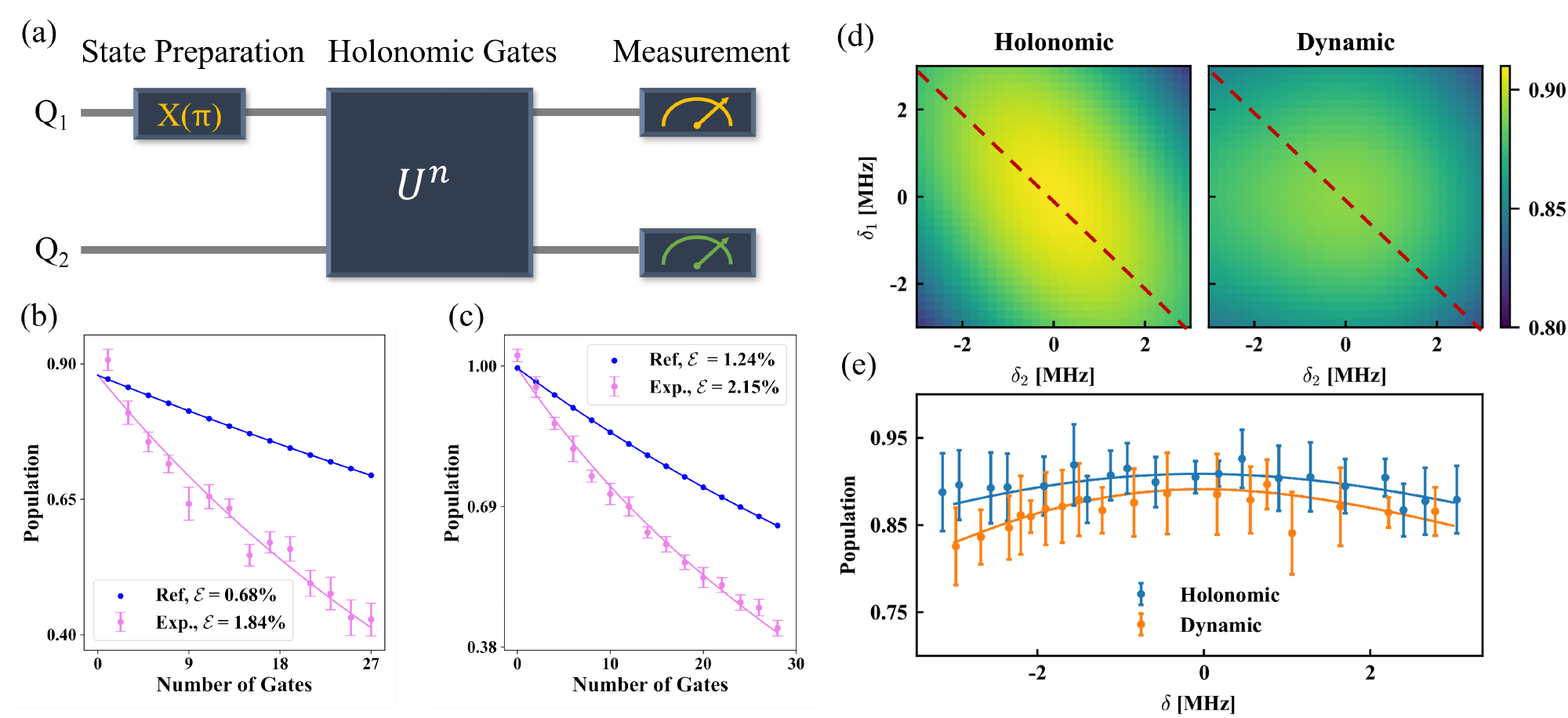}
    \caption{{\bf Error and Robustness feature of holonomic gates.} (a) Time sequence for the error estimation. We applied a series of SWAP ($\sqrt{\text{SWAP}}$) gates on the system and then measured the population of the qubit. (b)-(c) Result of the SWAP and $\sqrt{\text{SWAP}}$ gate. The blue solid line represents the fitted result of the numerical simulation that takes energy dissipation into account, while the purple solid line corresponds to the fitted result of the experimental data. After deducing the energy dissipation effect, the SWAP ($\sqrt{\text{SWAP}}$) average gate error can reach 1.16\% (0.91\%). (d) Robustness simulation results of the detuning-population relation considering decoherence. $\delta_1$ and $\delta_2$ refer to the flux pulse detuning of the two qubits. The left (right) panel represented to the holonomic (dynamic) SWAP gate. (e) The solid lines are the population corresponding to the red dashed lines in (d). At a detuning of 3 MHz, the dynamic SWAP gate exhibits 6.13\% population loss, while our scheme results in merely 3.46\% loss. The blue and orange dots (solid lines) are the experiment results (numerical simulations).}
    \label{fig:FiG3}
\end{figure*}

    The universal two-qubit gates set for quantum network nodes can be constructed by concatenating $\sqrt{\text{SWAP}}$ gates with single-qubit operations. Entanglement property can also be investigated by applying $\sqrt{\text{SWAP}}$ gate by adjusting $\tilde{g}_{12}(t)/\tilde{g}_{22}(t)=0.414$, 
  with envelope average of parametric modulation were engineered as $g^a_{12}/2\pi=4.16\text{MHz}$ and $g^a_{22}/2\pi=10.04\text{MHz}$.
  As demonstrated in Fig.\ref{fig:FiG2}(c), the Bell state $\frac{1}{\sqrt{2}}(|10\rangle+|01\rangle)$ was correctly built at 46 ns, with theoretical simulations predicting negligible population in cable mode $M_2$.
  To comprehensively characterize entanglement property, single qubit tomography and subspace tomography were performed, as shown in Fig.\ref{fig:FiG2}(d).
  For subspace tomography, we only focus on the subspace $\{|10\rangle,|01\rangle\}$. 
  The off-diagonal elements explicitly demonstrated the entanglement property of our quantum system. In our experiment, the final state fidelity $\mathcal{F} = Tr(\sqrt{\sqrt{\rho}\sigma\sqrt{\rho}})$ achieved $97.54\pm0.42\%$ after considering the decoherence effect in our simulation result, demonstrating the high-efficiency entanglement generation capability of the proposed protocol. 
  
  \textbf{Error Rate and Robustness.} We further characterized the error rates of our holonomic gates. By successively applying multiple SWAP and $\sqrt{\text{SWAP}}$ gates to the quantum system initialized in state $|10\rangle$, we measured the population dynamics of both qubits. Using a linear model to extract gate errors, we obtained an error rate $\mathcal{E} = 1.84\%$ for the SWAP gate and $\mathcal{E} = 2.15\%$ for the $\sqrt{\text{SWAP}}$ gate, as shown in Fig.\ref{fig:FiG3}(b) and Fig.\ref{fig:FiG3}(c). Notably, the $T_1$ relaxation and cavity mode leakage contributions were non-negligible. After compensating for energy dissipation effects, the SWAP($\sqrt{\text{SWAP}}$) gate error rates can be reduced to 1.16\% (0.91\%).

  The robustness of our scheme inherits the robustness advantage of geometric gates, which is validated theoretically and experimentally, as illustrated in Fig.\ref{fig:FiG3}(d) and Fig.\ref{fig:FiG3}(e). Here, dynamic gates refer to two parametric modulations $\pi$ gates applying on $Q_1$-$M_2$ and $Q_2$-$M_2$. $\delta_1$ and $\delta_2$ denote detunings induced by periodic pulses, with optimal SWAP gate fidelity achieved at $\delta_1=\delta_2=0$. To evaluate robustness, we introduce a frequency shift in mode $M_2$, simulating a systematic error that generates correlated detunings $\delta_1= -\delta_2= \delta$. The experimental data shows that our protocol is more resistant to system error than the dynamical pulse method. With 3 MHz detuning, the dynamic SWAP gate exhibits $R_d = 6.13\%$ population loss(orange curve). In contrast, our scheme demonstrates significantly reduced loss at $R_h = 3.46\%$(blue curve), with a relative improvement ratio $R_r = (R_d-R_h)/R_d = 43.6\%$, highlighting its enhanced robustness.

\begin{figure*}[!ht]
    \centering
    \includegraphics[width=18cm]{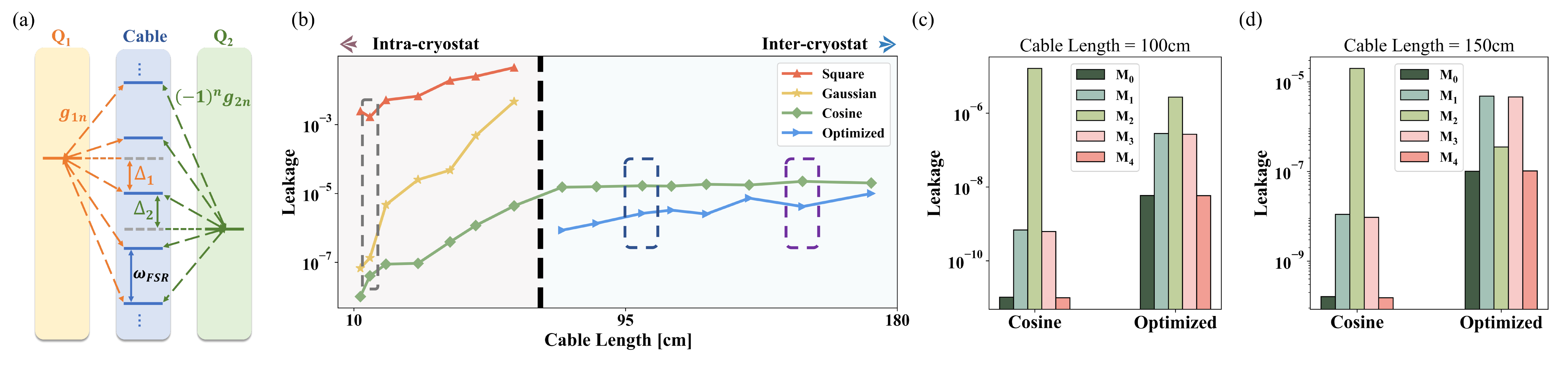}
    \caption{{\bf Waveform Optimization and Leakage Suppression.}  (a) Illustration of the coupling between qubits and cable modes. We considered the nearest five cable modes. The coupling strengths $g_{1n}, (-1)^ng_{2n}$ and detunings $\Delta_1, \Delta_2$ between the qubits and each mode are labeled. (b) The cable mode leakage of square, Gaussian, cosine and Adam optimized waveforms at different cable lengths, represented by the red, yellow, green and blue solid lines, respectively. The cable length ranges from 12cm to 172cm, corresponding to the FSR from 35MHz to 500MHz. Left (Right) panel represents the in-refrigerator (refrigerator-to-refrigerator) connection condition. The cable length (FSR) used in our experiments is approximately 15cm (403MHz), which is indicated by the gray dashed box. Adaptive Moment Estimation (Adam) was applied to the chosen guessed waveform to suppress cable mode leakage.  The application of Adam led to a suppression of leakage by one order of magnitude, demonstrating its effectiveness in improving waveform performance. 1m (FSR = 60MHz) and 1.5m (FSR = 40MHz) were represented by blue and purple dashed box, respectively. (c)-(d) Leakage Distribution Analysis. The leakage distribution was analyzed for cable length of 1m and 1.5m. At both lengths, the leakage in the target mode ($M_2$) was effectively suppressed. A slight increase in leakage was observed in the neighboring modes; however, this effect was minor and did not noticeably degrade the overall system performance.}
    \label{fig:FiG4}
\end{figure*}

\textbf{Discussion.} We demonstrate the implementation of remote two-qubit gates utilizing the holonomic quantum gate scheme. Experimental results show that the proposed scheme achieves gate operations in about 30 ns, exhibiting a speed advantage over existing remote connectivity schemes while maintaining inherent robustness. Systematic investigation of leakage dynamics under various waveforms reveals that leakage errors can be numerically suppressed to the $10^{-8}$ level in our sample (See Methods). We note that this result can be experimentally achieved by optimizing the experimental setup and reducing decoherence. Our work establishes a scalable framework for realizing remote coupling in a superconducting platform, which further provides essential components for quantum error correction, hybrid quantum computing and quantum networks.

\textbf{Waveform Optimization and Leakage Suppression.}  
Suppressing unwanted leakage to the cable modes is essential and critical for this cable-coupled architecture. We use numerical simulation to select the optimal frequency parameters and waveform shape\cite{werninghaus_leakage_2021,motzoi_simple_2009,gambetta_analytic_2011,chen_measuring_2016,rebentrost_optimal_2009}. We investigated the leakage suppression capabilities by performing numerical simulations with three pulse waveforms. The numerical simulations indicate that the cosine shape performs best between the three waveforms. For a more straightforward analysis, the Hamiltonian was transformed into the rotating frame, yielding the following form: 
\begin{equation}
        \begin{split}
        H/{\hbar}&= \sum_{j=0,1,2,3,4} g_{1j} B_1|M_j\rangle \langle Q_1|\\
        &+\sum_{j=0,1,2,3,4} g_{2j} B_2|M_j\rangle \langle Q_2| \\
            &+\sum_{j=0,1,3,4} (\omega_{M_j}-\omega_{M_2})|M_j\rangle\langle M_j|+ \text{H.c.} ,
        \end{split}
\end{equation}
where $B_{1(2)}=e^{-i(\omega_{Q_{1(2)}}-\omega_{M_2})t-iF_{1(2)}(t)}$, $F_{1(2)}(t)=A_{1(2)}(t)\sin[(\omega_{M_2}-\omega_{Q_{1(2)}}+\delta_{1(2)})t+\phi_{1(2)}]$. 
Using the Jacobi-Anger expansion, it can be obtained $B_{1(2)}=J_1[A_{1(2)}(t)/(\omega_{M_2}-\omega_{Q_{1(2)}})]e^{-i(\delta_{1(2)} t+\phi_{1(2)})}+J_2[A_{1(2)}(t)/(\omega_{M_2}-\omega_{Q_{1(2)}})]e^{-i[(\omega_{M_2}-\omega_{Q_{1(2)}}+2\delta_{1(2)})t+2\phi_{1(2)}]}$, where $J_1$ and $J_2$ are the first and second order Bessel functions of the first kind.
 As illustrated in Fig.\ref{fig:FiG4}(a), our simulation model accounted for five cable modes coupled to both qubits. For a fixed FSR, we varied the frequencies of the two qubits, ranging from $\omega_{M_2}$ to $\omega_{M_2} + \omega_{FSR}$ and from $\omega_{M_2} - \omega_{FSR}$ to $\omega_{M_2}$ respectively, with a step of 0.025 MHz. Through the optimization of $\omega_{Q_1}$ and $\omega_{Q_2}$, we minimized the leakage to the unwanted cable modes during this process. For the FSR in the experiment of approximately 400 MHz, the leakage was $1.65\times 10^{-3}$ for the square waveform, $1.34\times 10^{-7}$ for the Gaussian waveform, and $4.06\times 10^{-8}$ for the cosine waveform. These results confirm that the joint optimization of qubit frequencies and pulse waveforms provides dual pathways for fidelity enhancement in future implementations. In the laboratory frame, numerical simulations indicate that the leakage can be effectively suppressed to the order of $10^{-3}$ using a cosine waveform with an optimized frequency, which is well beyond our readout fidelity.

Additionally, we characterized leakage dependence on FSRs of our scheme, as shown in the left panel of Fig.\ref{fig:FiG4}(b).  Since the length of the cable is inversely proportional to its FSR, as we gradually increase the cable length, the problem of leakage to other modes due to the decrease of FSR will become more severe. For distributed quantum computing within a dilution refrigerator, a connection length of 0.6 m is sufficient, corresponding to a FSR of 100 MHz.. As illustrated in the figure, the leakage values for the cosine waveform is $4.33\times 10^{-6}$, underscoring its suitability for facilitating distributed superconducting quantum computation. This characteristic aligns well with the long-distance coupling requirements within the dilution refrigerator, making it a promising candidate for such applications.

For refrigerator-to-refrigerator connections, a smaller FSR significantly increases cable mode leakage. To mitigate this effect, the Adaptive Moment Estimation (Adam) optimizer can be employed to suppress cable mode leakage effectively. With its adaptive learning rate feature, the Adam optimizer effectively alleviates the issue of local optima, as illustrated in Fig. \ref{fig:FiG1}(d).The loss function is defined as the difference between the ideal and simulated evolution matrices, resulting in the form $\text{Loss} = 1 - (\left|\text{Tr}(V^\dagger U)\right|/d)^2$, where $V$ denotes the ideal evolution matrix, $U$ represents the simulated evolution matrix, and $d$ is the dimension of the evolution matrix. Specifically, we focused on the regime of very small FSR values, and numerical result was illustrated in the right panel of Fig.\ref{fig:FiG4}(b), Fig.\ref{fig:FiG4}(c) and  Fig.\ref{fig:FiG4}(d). For an FSR of 40 MHz, corresponding to a 1.5-meter cable length, the cable mode leakage was suppressed to $4.15\times 10^{-6}$, achieving an improvement of one order of magnitude compared to the conventional cosine waveform approach. Although the Adam method induced a slight increase of neighboring modes leakage, the overall cable mode leakage was significantly suppressed, demonstrating the effectiveness of the Adam optimizer in improving performance under conditions of smaller FSR.

\bibliographystyle{modified-apsrev4-2}
\bibliography{refs}

\textbf{Author Contribution.} X.T. designed the scheme, and Y.L., X.Z., Y.Z. and R.G. performed it. Y.L. and X.Z. analysed the data.  W.Z. designed and fabricated the sample. Y.L., X.Z. and X.T. wrote the manuscript. Y.Y. and X.T. supervised the project.

\begin{acknowledgements}
\textbf{Acknowledgements.} This work was partly supported by the Innovation Program for Quantum Science and Technology (2021ZD0301700). NSFC (Grant No. 11890704, and No. U21A20436), NSF of Jiangsu Province (Grant No. BE2021015-1, BK20232002), and 
Natural Science Foundation of Shandong Province (Grant No. ZR2023LZH002).
\end{acknowledgements}

\end{document}